\documentclass[12pt]{article}
\usepackage[english]{babel}
\usepackage[cp866]{inputenc}
\usepackage{amssymb}
\usepackage{amsmath}
\usepackage{amsfonts}

\usepackage{graphicx} 

\usepackage{xcolor}

\newcommand{\blue}[1]{\textcolor{blue}{#1}}

\begin{document}
\thispagestyle{empty}

\author{\normalsize\bf K.A. Bronnikov, M.I. Kalinin, and V.V. Khruschov\\[3pt]
			{\small\it
			Center for Gravitation and Fundamental Metrology, VNIIMS, Moscow, Russia }}
			
\title{\Large\bf On the heat evolution of the early Universe}
\date{}
\maketitle

\begin{abstract}
\noindent
   We discuss the heat history of the early Universe and its further evolution in the framework 
   of modern cosmological models of general relativity (GR) and alternative theories of
   gravity. Of great importance are the new puzzling forms of matter comprising parts of our
   Universe, namely, dark energy and dark matter that crucially affect the Universe structure 
   and evolution.

\end{abstract}

\section*{\large Introduction}

  Measurements are an element of almost any human activity. With the beginning of the
  exploration and development of outer space, it became especially necessary to carry out
  astronomical, astrophysical and cosmological measurements. It is amazing that hypothetical
  theories on the structure and development of our whole Universe can now be tested using the 
  results of measurements carried out both on Earth and in outer space.

  This paper briefly discusses the thermal history of the early Universe and the associated
  development of some modern cosmological models using general relativity (GR) and 
  alternative theories of gravity. Attention is drawn to the decisive role of new mysterious 
  forms of matter, namely, dark energy and dark matter, in the evolution and structure of the 
  Universe. The main markers in this case are the readings of temperature measurements
  of the cosmic relic radiation (Cosmic Microwave Background, CMB), either carried out 
  or supposed ones, which are a kind of Ariadne's thread in the tangled labyrinth of models 
  of the Universe.

  The outline of this paper is as follows. Section 1 briefly discusses the emergence history
  of modern ideas on the structure and evolution of the early Universe. The main role here 
  is played by the concept of inflation --- the ultra-fast expansion of the Universe at a very 
  early stage of its development --- and the Big Bang, which formed the currently observed
  Universe. Section 2 discusses the baryogenesis phenomenon and the radiation-dominated 
  epoch. For the thermal history of the Universe, this epoch is of primary interest. Section 3 
  describes the characteristics of primary nucleosynthesis and the transition to recombination. 
  Section 4 describes the formation of CMB, the background radiation whose temperature 
  characteristics are the basic criteria for the applicability of modern cosmological models. 
  Section 5 is a conclusion, where some issues that  require resolution in the future are 
  discussed, in particular, those related to the possible variability of fundamental physical 
  constants.

\section{\large  The emergence of modern ideas about the early Universe}

  At the beginning of the 20th century, the Universe seemed to be a static formation,
  consisting of a large number of stars moving according to the laws of Newtonian
  mechanics and Newton's theory of gravity. After creation of the special theory of 
  relativity (SR), which formulated the laws of mechanics of large relative velocities 
  and changed the Newtonian idea of space and time, Albert Einstein set the task of 
  describing gravity taking into account the laws of SR, which resulted in creation of 
  the general theory of relativity (GR) \cite{Einstein}. The equations of this theory 
  described the motion of celestial bodies at any time and location, but it did not seem 
  possible to obtain a stationary solution for the Universe as a whole. Trying to preserve 
  the possibility of a stationary state of the Universe when describing it in the framework 
  of GR, Einstein introduced the so-called $\Lambda$-term (cosmological constant) to 
  the GR equations. Some years later, Alexander Friedmann showed that, in the general 
  case, cosmological solutions of Einstein's equations (including those with a cosmological 
  constant) are nonstationary, and the Universe should expand or contract over time 
  \cite{Friedmann}. In 1927, George Lema\^{\i}tre independently obtained Friedmann's
  solutions and actually introduced the concept of the Big Bang of the Universe, calling 
  it at that time ``the primary atom'' \cite{Lemetre}. Einstein's static model is a special
  solution to the equations of GR, and it was also shown to be unstable. After the 
  discovery of the real expansion of the Universe \cite{Hubble}, Einstein claimed that 
  the introduction of a cosmological constant had been his mistake. Nevertheless, in 
  the modern theory of gravity, the cosmological constant plays an important role and 
  is used, for example, for a description of the inflationary periods in the development 
  of the Universe.

  Nonstationary isotropic cosmological solutions of Einstein's equations, similar to 
  Friedmann's, are now used to study various periods of the Universe evolution, 
  starting from a singularity (an initial state with an infinite density of matter) 
  and ending with the modern stage of accelerated expansion of the Universe.

  Soon after the theoretical predictions of Friedmann and Lema\^{\i}tre, the results 
  of astronomical observations of Cepheids in spiral nebulae were obtained, and they
  allowed Edwin Hubble to propose his law of the proportionality between the 
  redshift of stellar radiation in spiral nebulae and the distance from the Earth 
  \cite{Hubble}. The proportionality coefficient was named the Hubble constant $H$, 
  and the redshift was interpreted as a manifestation of the recession rate of the
  ``nebulae'' (as is now understood, galaxies similar to ours), and a conclusion 
  was made on the expansion of the Universe as a whole, which leads to the 
  recession of galaxies.

  The nature of the cosmological constant $\Lambda$ still remains a mystery. 
  Erast Gliner in the late 1960s suggested that it is related to the physics of vacuum, 
  that is, a medium without particles. This interpretation is confirmed by the fact that
  the physical properties of the medium described by the $\Lambda$-term do not 
  depend on the reference frame, which is impossible for any substance or particles --- 
  they always have a preferred reference frame in which the substance is at rest. In modern 
  cosmology, the $\Lambda$-term is used in two ways: on the one hand, as the driving
  force of the initial inflation and, on the other hand, as ``dark energy'' --- the cause of 
  modern accelerated expansion of the Universe. In these two applications, the values 
  of the constant $\Lambda$ differ by many orders of magnitude, which is one of the 
  manifestations of the so-called ``cosmological constant problem,'' which has not yet 
  been solved. A possible solution to this problem is to introduce a ``cosmological 
  constant'' depending on cosmic time in the form of the potential of some scalar or 
  vector field.

  Before discussing the thermal history of the Universe, it makes sense to list the 
  main stages of its evolution in accordance with modern ideas.

\paragraph{1.1. Initial (pre-inflationary) stage.}
  To date, there are no definite or dominant ideas on the initial stage of the emergence
  of the Universe, obeying quantum laws for both the gravitational field and matter. 
  A semiclassical formulation, with classical gravitational field equations and quantum
  equations for matter, is often used. Solutions to the classical equations of gravity in 
  most of the theories include an initial state in the form of a singularity with infinite
  space-time curvature and infinite density of matter, which cannot be regarded as an
  acceptable concept, but indicates the inability of such theories to adequately describe 
  extremely large curvatures and/or densities. A number of theories suggest a transition 
  of the Universe from compression to expansion through a regular ``bounce,'' but 
  the most promising point of view seems to be the emergence of the classical Universe
  from a fluctuation of ``quantum space-time foam'' or some quantum vacuum state 
  (see, for example, \cite{Soo}). One of the first models of such a transition has been
  presented in \cite{Kalinin}.

\paragraph{1.2. Inflation,} a stage of ultrafast expansion of the Universe, with duration 
  $t$ within $\sim 10^{-32}$ s. The idea of inflation makes it possible to solve a number
  of difficult problems in cosmology, caused by the absence of a causal connection 
  between distant parts of the Universe, which makes inexplicable their identical properties.
  Due to inflation, the entire observable region of the Universe could arise from a small 
  causally connected region of space. Some alternative theories, instead of inflation, assume 
  a nonsingular ``bounce'' of the Universe after its compression (collapse). And the 
  existence of this stage then explains the causal connection between distant regions 
  of the Universe. Inflation was first considered on the basis of Einstein's equations with 
  a cosmological constant and a zero energy-momentum tensor, when the Universe 
  begins to develop according to de Sitter's inflationary solution. The work of Alexei 
  Starobinsky \cite{Starobinsky} played an important role in this inflation scenario. 
  Then significant progress in the theory of inflation was achieved in the works of 
  Alan Guth and Andrei Linde \cite{Guth, Linde1, Linde2}.

\paragraph{1.3. Post-inflationary heating.} 
  The birth of matter from vacuum ($10^{-32}\, {\textrm s} < t < 10^{-5}\, {\textrm s}$, 
  hereinafter time is counted from the beginning of inflation) is the stage of ``exit'' from
  inflation and transformation of vacuum energy into the energy of matter and radiation.  
  Matter in this case consists of elementary components such as quarks and leptons, 
  and the radiation consists of gauge bosons. At the beginning of this stage, a 
  unification of strong, electromagnetic and weak interactions takes place, then occurs 
  violation of this unification (the strong interaction is separated), confinement of 
  quarks (capture of quarks, antiquarks and gluons) with the formation of baryons 
  and mesons, and baryogenesis.

\paragraph{1.4. The radiation dominated era.}
  The ``hot Universe'', which takes a period from $10^{-5}\, {\textrm s}$ to 
  $10^{12}\, {\textrm s} \sim 30000\, \textrm{years}$ and is characterized by the 
  fact that it is radiation and ultrarelativistic particles that make the main contribution 
  to the energy density of the Universe. It is during this epoch that the main processes
  in the history of matter we know (baryogenesis and primary nucleosynthesis) and 
  the thermal history of the Universe took place, to be discussed below in this article.

\paragraph{1.5. The matter dominated era.}
  Since with the expansion of the Universe, the radiation energy density 
  ($\sim a^{-4}$, where $a = a(t)$ is the scale factor of the Universe growing with time) 
  falls faster than the energy density of nonrelativistic matter ($\sim a^{-3}$), from a 
  certain time instant ($t \sim 3\times 10^4\, \textrm{years}$), matter becomes dominant 
  and remains so for a long time, up to $t \sim 10^{10}\, \textrm{years}$. At the 
  same time, not all nonrelativistic matter is observed in the form of stars, nebulae, etc.: 
  its main part (about 80\% --- the so-called dark matter) remains invisible and 
  manifests itself in observations only due to its gravitational field.

\paragraph{1.6. The modern era of dark energy dominance.} 
  It starts at $t \sim 10^{10}\, \textrm{years}$, or about $3.7\times 10^9$ years 
  ago, corresponding to a cosmological redshift $z \sim 0.4$. According to the available 
  observational data, in the modern era, the expansion of the Universe is accelerating, 
  and this, according to GR, is possible only due to the influence of some unknown 
  source of the gravitational field, close in properties to the cosmological constant 
  $\Lambda$, which is called dark energy (DE). In addition to the model with a 
  $\Lambda$-term, there are many models within the framework of GR that identify 
  DE with various physical fields. An alternative approach is to abandon GR and use 
  other, ``generalized'' theories of gravity, where cosmological solutions describe 
  the accelerated expansion of the Universe without invoking unknown sources.

  Note that the thermal history of the Universe begins from the third stage, when it 
  already makes sense to introduce the concept of temperature as the average value 
  of energy (minus the rest energy) per particle. In what follows, we will use both 
  energy units (eV, GeV, etc.) and kelvins for temperature on equal grounds;
  1 eV corresponds to approximately 11604 K.

\section{\large Baryogenesis and the radiation dominated era}

  {\it Baryogenesis}, that is, the process of emergence of an excess of matter in 
  the Universe as compared to antimatter, begins before the spontaneous violation of 
  gauge electroweak symmetry (the scale of the average energy or temperature of 
  such violation is about 100 GeV) and ends after the post-inflationary heating, when 
  quark and gluon confinement occurred, i.e. at a temperature of about 1 GeV. 
  At a temperature of the order of 100 MeV, the radiation epoch begins, in which 
  a transition from the early Universe to the modern one gradually happens, and the
  thermal characteristics of, for example, the cosmic microwave background radiation 
  can be measured in the modern epoch with high accuracy 
  \cite{Aghanim1, Aghanim2, Abdullah}.

  It is generally supposed that the processes occurring in the early Universe (above all, 
  the rapidly changing gravitational field) lead to the birth and subsequent annihilation 
  of matter and antimatter. However, in the modern Universe, we observe mostly matter 
  consisting of nuclei and electrons. The nuclei mainly consist of baryons, and these 
  are mostly protons and neutrons (besides, there are baryon resonances that enter 
  into excited states of nuclei). The causes and course of the occurrence of matter 
  excess over animatter (BAU --- Baryon Asymmetry of the Universe) are not yet fully 
  established. There are a number of models for the emergence of BAU in the process 
  of so-called baryogenesis (see, for example,  \cite{Dine}). Generally, the basis of such
  models includes the following conditions formulated by Andrei Sakharov
  \cite{Sakharov}: (1) violation of baryon number conservation; (2) violation of C- and 
  CP-symmetries;\footnote{C-symmetry is a symmetry with respect to the replacement 
	  	of particles by antiparticles and vice versa, CP-symmetry is a combination of 
  		spatial reflection and C-symmetry} 
  (3) violation of the thermal equilibrium of ongoing reactions. 
  (C-symmetry is an exact symmetry between the properties of particles and 
  antiparticles, and CP-symmetry assumes, in addition, symmetry under reflections 
  against a plane in space.)  For example, during phase transitions, baryon asymmetry 
  can arise around the expanding walls of bubbles, whose surfaces form boundaries 
  between different phases. Another example assumes particle decays that take place 
  when the thermal equilibrium is disturbed.

  The existing baryon asymmetry of the Universe is usually characterized by the 
  parameter $\eta_{\rm B}$, whose value may be extracted from observational 
  results \cite{Aghanim2}:
$$
		\eta_{\rm B} = 
		\frac{n_{\rm B}-n_{\bar{\rm  B}}}{n_{\rm B}+ n_{\bar{\rm  B}}}
		\sim \frac{n_{\rm B}-n_{\bar{\rm  B}}}{n_{\gamma}} 
		\sim \frac{n_{{\rm B}_0}}{n_{\gamma}} 
		\sim 6\times 10^{-10},
$$
  where  $n_{\rm B}$ and $n_{\bar {\rm B}}$ are, respectively, the baryon number 
  density and the antibaryon number density by the end of the radiation epoch, 
  $n_{{\rm B}_0}$ is the modern baryon number density that remained due to the 
  excess of the number of baryons over the number of antibaryons in the early epoch, 
  $n_{\gamma}$ is the number density of photons in the CMB radiation. Such 
  qualitative estimates arise under the assumption that a substantial part of the relic 
  photons emerged as a result of annihilation of primary baryon-antibaryon pairs, which
  had formed when there was already an asymmetry between baryons and antibaryons.   
  There is no generally accepted theory of BAU yet; the most common are models 
  based on electroweak baryogenesis \cite{Kuzmin}.

{\it The radiation dominated era\/} is one of the most important periods in the development 
  of the Universe. The main contribution to the energy density in this epoch comes 
  from relativistic particles, i.e., photons, neutrinos and fast electrons. By temperature, 
  this era occupies the range from approximately 100 MeV to 1 eV, and its lifetime 
  lasts from $10^{-5}$ s to $10^{12}$ s (approximately $3\times 10^4$ years). 
  During a period near $t \sim 1$ s, at a temperature of about 2 MeV, neutrinos leave 
  the thermal equilibrium with the plasma and begin to propagate freely in the Universe. 
  Almost immediately after that, electrons and positrons become nonrelativistic and
  annihilate, transferring their energy to photons. At the same temperature of about 
  2 MeV, the {\it primary nucleosynthesis} begins, which continues up to temperatures of
  the order of 0.03 MeV. Soon after the end of the {\it radiation dominated era}, at
  temperatures of the order of 1 eV, occur the last acts of photon scattering inside 
  the plasma (at $T\sim 0.3$ eV), and the {\it cosmic microwave background radiation}    
  (CMB, Cosmic Microwave Background) is finally formed.

\section{\large  Primary nucleosynthesis, recombination and CMB formation}

  The {\it primary nucleosynthesis} (Big Bang Nucleosynthesis, BBN) begins when 
  the photon gas has been sufficiently cooled, so that the simplest nuclei could form 
  and exist. The gas temperature at the beginning of BBN is about 2 MeV, and at the 
  end it is about 0.03 MeV. The total time of primary nucleosynthesis lasts from 
  $t\sim  1$ s to $t\sim \blue{10^2 \div 10^3}$ s. During this period, the simplest 
  nuclei are formed, 
  mainly including the nuclei of hydrogen (H), deuterium (D, 2H), tritium (T, $^3$H), 
  helium-3 ($^3$He), helium-4 ($^4$He), lithium-7 ($^7$Li) and beryllium-7 ($^7$Be).   
  Protons and neutrons have sufficient kinetic energy to form deuterium nuclei, which 
  can further participate in nuclear reactions without decaying under the impact of
  photons. To solve a complex system of nonlinear equations describing nuclear 
  reactions, numerical calculation programs were created that led to the abundance 
  values for the simplest elements (with the exception of $^7$Li) coinciding with those 
  observed in nature. Nuclear reaction rates depend on the temperature and the 
  baryon number density, so the ratio of the baryon number density to the photon 
  number density plays an important role.

  When the temperature of the plasma, consisting of nuclei, electrons and photons, 
  drops to a sufficiently low level of the order of 0.3 eV, there starts recombination of  
  electrons and nuclei into neutral atoms, and the plasma becomes transparent to 
  photons, which can now freely propagate throughout the Universe, forming the 
  now observed Cosmic Microwave Background. The existence of the CMB was 
  predicted by George Gamow within the framework of his hot Universe model 
  \cite{Gamow1, Alfer, Gamow2}.

\section{\large The temperature characteristics of the early Universe and the CMB}

  When discussing the thermal history of the early Universe, of great importance are
  the reference points characterizing the formation of a vacuum condensate which 
  breaks the unification symmetry of strong, weak and electromagnetic interactions 
  on a scale of the order of $10^{16}$ GeV, and the formation of another vacuum
  condensate which breaks the unification symmetry of electroweak (electomagneic and
  weak) interactions on a scale of the order of $10^2$ GeV. The confinement of quarks 
  and gluons occurs on a scale of about 1 GeV.

  As already mentioned, neutrinos separate from the primary plasma much earlier 
  than photons, at $t\sim 1$ s and at a temperature of about 2 MeV. The further 
  evolution of the neutrino gas occurs independently from the CMB; its temperature 
  differs from the CMB temperature and is currently $1.96 \pm 0.02$ K, in agreement 
  with theoretical predictions \cite{Follin}.

  The CMB has nowadays a temperature of $2.7255 \pm 0.0006$ K within standard
  uncertainty \cite{Workman} and has an almost isotropic 
  structure (minus the dipole component, mainly associated with the motion of the 
  Solar System). Modern precision measurements of the CMB temperature have made 
  it possible to detect inhomogeneities at a level of $10^{-5}$, which correspond 
  to inhomogeneities in the recombination era that have led, in the process of further 
  evolution, to the formation of the large-scale structure of the present-day Universe.

\section{Conclusion}

  To reconstruct the Universe's history, one usually invokes GR and the recently 
  obtained data of observational cosmology, primarily from the Planck space telescope
  \cite{Aghanim1, Aghanim2}. However, it should be noted that a number of currently 
  existing scenarios are hypothetical. For example, to obtain quantitative estimates, 
  the known values of fundamental physical constants (FPCs) are used. But since the 
  stages of the Universe evolution occupy cosmic periods of time, during these periods, 
  the values of the FPCs could change, which then inevitably affects the characteristics 
  of the early Universe evolution. Refs. \cite{Martins, Bronnikov} considered the restrictions
  obtained to date on long-term variations of the gravitational constant, the fine structure
  constant, and the proton-to-electron mass ratio. The strongest limits on variations of 
  the fine structure constant and the proton-to-electron mass ratio were obtained in 
  an experiment comparing atomic clocks \cite{Lange}. The theoretical dependences 
  of nuclear reaction rates on the fine structure constant are given in \cite{Meissner}. 
  The values of the fine structure constant and the proton-to-electron mass ratio underlie 
  the interpretation of atomic and molecular spectra of celestial bodies. Therefore, 
  their possible dependence and the dependence of the gravitational constant on 
  cosmic time are of fundamental importance in obtaining quantitative characteristics of 
  the early Universe. However, the presently existing models of the early Universe are facing,
  among other challenges, the problem that consists in different values of the Hubble 
  constant, determined from the recession of supernovae and from the CMB 
  characteristics (see, for example, \cite{Vagnozzi}). Thus the theory of the Universe
  evolution is still far from being completed.


\end{document}